\documentclass[]{raa}       % referee 
\usepackage{graphicx,times}
\usepackage{natbib}
\usepackage{amssymb,amsmath}
\bibpunct{(}{)}{;}{a}{}{,}

\usepackage[a4paper=true,pagebackref=true]{hyperref}
\hypersetup{pdftitle = 3D photometric modelling of galaxies, pdfauthor = E. Tempel et al., pdfsubject= , pdfkeywords = keyword1 keyword2 keyword3} 
\hypersetup{colorlinks = true, linkcolor = green, anchorcolor = red, citecolor = blue, filecolor = red, pagecolor = red, urlcolor = red}

\begin{document}

   \title{Recovering 3D structural properties of galaxies from SDSS-like photometry}

 \volnopage{ {\bf 2015} Vol.\ {\bf 15} No. {\bf 10}, 1613--1628}
   \setcounter{page}{1}

   \author{E. Tempel\inst{1}, A. Tamm\inst{1}, R. Kipper\inst{1,2},  P. Tenjes\inst{1,2}
   }
%% Here is an example of three authors come from different institutes.
%% For single author or all the authors from an institute, use "\inst{}" only

   \institute{ Tartu Observatory, Observatooriumi~1, 61602 T\~oravere, Estonia; {\it elmo.tempel@to.ee}\\
%% Please give the E-mail address of the author, to whom future correspondence and
%% offprint requests will be sent.
        \and
             Institute of Physics, University of Tartu, Ravila 14c, 50411 Tartu, Estonia\\
\vs \no
   {\small Received 2015 January 12; accepted 2015 March 13}
}

\abstract{Because of the 3D nature of galaxies, an algorithm for constructing spatial density distribution models of galaxies on the basis of galaxy images has many advantages over surface density distribution approximations.
We present a method for deriving spatial structure and overall parameters of galaxies from images and estimate its accuracy and derived parameter degeneracies on a sample of idealised model galaxies. The test galaxies consist of a disc-like component and a spheroidal component with varying proportions and properties. Both components are assumed to be axially symmetric and coplanar. We simulate these test galaxies as if observed in the SDSS project through $ugriz$ filters, thus gaining a set of realistically imperfect images of galaxies with known intrinsic properties. These artificial SDSS galaxies were thereafter remodelled by approximating the surface brightness distribution with a 2D projection of a bulge+disc spatial distribution model and the restored parameters were compared to the initial ones.
Down to the $r$-band limiting magnitude 18, errors of the restored integral luminosities and colour indices remain within 0.05~mag and errors of the luminosities of individual components within 0.2~mag. Accuracy of the restored bulge-to-disc ratios (B/D) is within 40\% in most cases, and becomes worse for galaxies with low B/D, but the general balance between bulges and discs is not shifted systematically. Assuming that the intrinsic disc axial ratio is $\le 0.3$, the inclination angles can be estimated with errors $<5^\circ$ for most of the galaxies with $\mathrm{B/D} < 2$ and with errors $<15^\circ$ up to $\mathrm{B/D} = 6$. Errors of the recovered sizes of the galactic components are below 10\% in most cases. The axial ratios and the shape parameter $N$ of Einasto's distribution (similar to the S\'{e}rsic index) are relatively inaccurate, but can provide statistical estimates for large samples. In general, models of disc components are more accurate than models of spheroidal components for geometrical reasons. 
}

   \authorrunning{E. Tempel et al. }            %author_head in even pages
   \titlerunning{Recovering the 3D structural properties of galaxies}  % title_head in odd pages
   \maketitle

%________________________________________________ sections below
%%======================================================================
\section{Introduction}
\label{sect:intro}

Huge galaxy imaging datasets like the Sloan Digital Sky Survey \citep{York:00} provide a good opportunity for giving statistically reliable estimates of galactic structural parameters, luminosities, colours, etc., e.g. to seek dependencies on the environment density or redshift. While such extensive galaxy samples help to reduce the noise caused by the cosmic variance, possible systematic errors resulting from specific parameter extraction procedures or imaging limitations may nevertheless introduce artificial trends or disguise the true ones. Thus it is important to test the consistency of the measured  parameters of galaxies at different distances, inclination angles, morphological types, colours, etc., before any general conclusions about the galaxies themselves can be drawn.

Most commonly, studies of galactic parameters are based on 2D surface brightness distribution approximations, derived with automated fitting tools GALFIT \citep{Peng:10}, GIM2d \citep{Simard:02}, BUDDA \citep{deSouza:04}, PyMorph \citep{Vikram:10}, DECA \citep{Mosenkov:14} etc. If handled correctly, these relatively fast and robust tools allow a quite reliable determination of most of the principal structural parameters of galaxies. However, much care is needed while interpreting 2D approximations of galaxies, since they are inconsistent by nature. For example, due to projection effects, the surface brightness and the S\'{e}rsic index of a galactic disc are considerably dependent on the inclination angle.

In reality, galaxies are 3D objects and for many applications, we are interested in the 3D properties of galactic structures: the actual shapes, inclination angles, alignments with other structures, etc. Under certain assumptions, these parameters can be derived by deprojecting the 2D surface brightness distributions. However, it is more straightforward to start by constructing a 3D model galaxy and determining its parameters by fitting its projection to the actual imaging data. Besides being more straightforward, 3D models have several advances over 2D ones. While 2D models of flattened galactic structures (most importantly, their central densities and density distributions) are inclination-dependent, 3D models are not. If one wishes to include (or study) the effects on intrinsic dust absorption, the inclusion of a dust disc into a 3D model would directly allow to consider dust extinction and reddening along each line of sight. Moreover, if kinematical data are available, one would be tempted to combine photometric and dynamical models, in which case a description of galaxies in terms of spatial densities would be needed. And finally, a flexible method for producing parametrised 3D galaxy structures and their 2D projections is very practical for testing the reliability of surface photometry techniques on realistic model galaxies.

Despite their apparent handiness, the derivation of the parameters of 3D galactic models from galaxy images can be a very degenerate exercise. In the case of a galaxy with triaxial ellipsoidal symmetry, free parameters are the two axial ratios and the two orientation angles of galaxy. In some cases, the degeneracy can be (partly) broken by extracting the inclination angle from additional information about the kinematics and the gas distribution \citep[see e.g.][]{Bertola91,Tenjes93, BakStat00, vandeven09, Kipper:12}. Unfortunately, the number of galaxies with the required additional data is not sufficient for statistics.

The 3D properties of galaxies can be deduced from the surface brightness distribution by assuming that galaxies have some additional symmetry in their overall mass distribution. Although the assumption of spherical symmetry reduces degeneracy completely this assumption is rather unrealistic for most of galaxies. A realistic and quite common assumption is to assume that the spatial density distribution of a galaxy is axisymmetric. In this case for disc-like galaxies, which can be described as a superposition of a flat disc component and a spheroidal bulge component, the inclination angle can be determined from the apparent flatness of the disc by assuming that the intrinsic axial ratio of the disc is small. Although the precise intrinsic flatness of the disc remains unknown, constraints can be laid using observations of edge-on galaxies \citep{Kautsch:06, Padilla:08, Rodriguez:13}. With reasonable accuracy, the derived disc inclination can be ascribed to the bulge as well. Although in principle, galactic components need not to be aligned, a study of a sample of 2MASS edge-on galaxies \citep{Mosenkov:10} indicates that in most cases, bulges and discs indeed are coplanar.

In apparently elliptical galaxies, extended disc-like structures can often be detected \citep{Krajnovic:08,Krajnovic:13, Arnold:14}. For example, 2D photometric studies of large samples of SDSS galaxies have demonstrated that the bulge-to-disc luminosity ratio (B/D) does not show a clear-cut separation between spheroidal and disc-like galaxies; instead, galaxies occupy the whole range of B/D \citep{Simard:11,Lackner:12,Mendel:14,Meert:15}\footnote{Note that the continuum in B/D from spheroidal to disc galaxies is partly caused by the automated modelling procedure, where adding a second component helps to reduce the $\chi^2$ value. However, for single component fits, the distribution of  S\'{e}rsic indexes is also continuous.}. Thus in addition to clearly disc-like galaxies, it should be feasible to estimate the inclination angles of many galaxies with rather large B/D ratios. However, the accuracy of such estimates has yet to be tested.

In the present paper we describe the construction and usage of an axisymmetric 3D galaxy model which can be fitted to single or multi-wavelength galaxy imaging. For testing the model in a realistic situation, we create a sample of simple idealistic 2-component (spheroid\,+\,disc) test galaxies. We plant these test galaxies into artificial Sloan Digital Sky Survey (SDSS) images and remodel them using our modelling software. By comparing the restored galactic parameters to the initial ones, we determine the limitations of the interplay between the SDSS imaging and the 3D modelling method.

%%======================================================================
\section{Creation of test galaxy images}
\label{sect:sdss_data}

%---------------------------------------------------------------------------
\begin{figure*}
  \centering
  \resizebox{0.50\hsize}{!}{\includegraphics{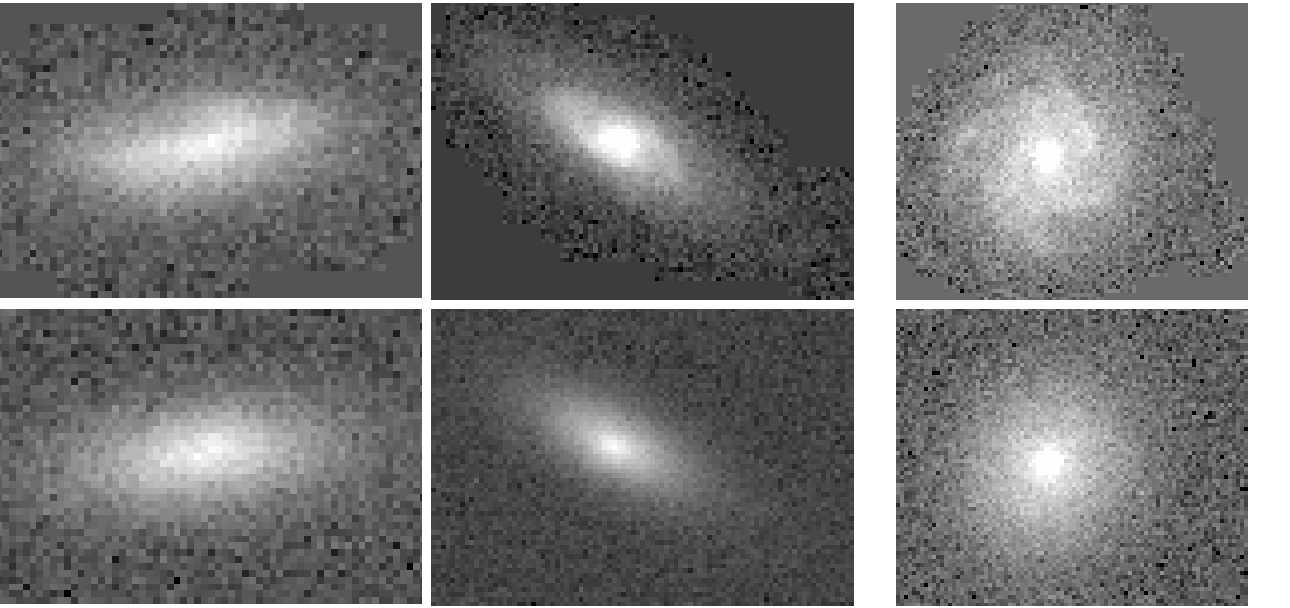}}
  \resizebox{0.49\hsize}{!}{\includegraphics{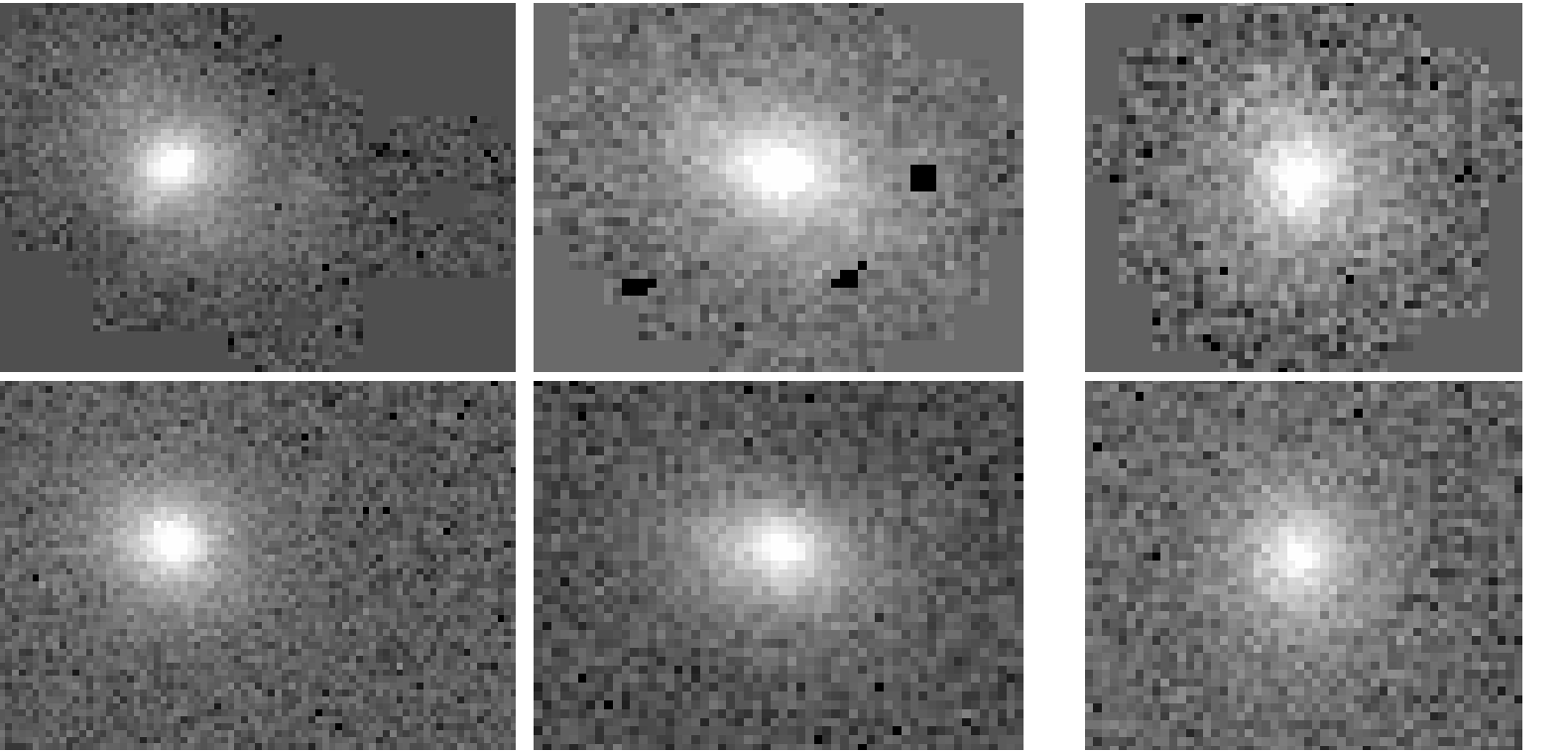}}
  \caption{Examples of the original (\emph{upper row}) and simulated (\emph{lower row}) SDSS images of the objects in our test galaxy sample. In the simulated images, the SDSS PSF and noise have been applied to the model galaxies (see text for more information). The smoothed and black areas in the original images are masked pixels.}
  \label{fig:mdl_sample}
\end{figure*}

To be able to give a correct estimate of the precision of a galaxy modelling technique, one needs a sufficiently large sample of test objects, for which the actual properties are well known. One would probably find that the best way to know the 3D properties of a galaxy is to construct the galaxy oneself.

We have constructed a sample of 1000 test galaxies on the basis of actual objects from the SDSS Data Release~8 \citep{Aihara:11}, covering a maximally broad range of the main properties: luminosities, luminosity distributions, sizes, axial ratios, B/D and inclination angles. Considering possible forthcoming applications of the model, galaxies with $r$-filter luminosities higher than 17.77~mag were selected from the original SDSS images, which is the completeness limit of the spectroscopic sample of the SDSS \citep{Strauss:02}.

The atlas (prefix `fpAtlas'), mask (prefix `fM'), and point-spread-function (prefix `psField') images of the observed galaxies were retrieved from the SDSS Data Archive Server (DAS). We used the SDSS software utilities \texttt{read\-Atlas\-Images} and \texttt{readPSF} for the atlas and PSF images, respectively. Masks were applied to the atlas images, using galaxy positions (given by \texttt{rowc} and \texttt{colc} in SDSS CAS). The model objects were created from the actual galaxy images using the same technique as described in Sect.~\ref{sect:model_simple}. At this step, the accuracy of the model is not crucial, since we would use only the modelled parameters further on. As a result we obtained 1000 idealised bulge+disc objects with a wide range of parameters. Note that in such models, the ``bulge'' actually represents all spheroidal components: bulges, stellar haloes, elliptical galaxies. Thus we can also have a ``pure bulge'' object, which is literally meaningless. However, we will keep using the term ``bulge'' instead of ``spheroid'' below for consistency and readability reasons, and for the applicability of the term ``bulge-to-disc ratio''.

Next we simulated these ideal model galaxies as if observed by the SDSS. The surface brightness distributions in $ugriz$ colours of the test galaxies were convolved with the SDSS point spread function (PSF). Poisson noise was added to every pixel according to the intensity value of the pixel. All other possible noise sources (the sum of sky foreground and background, the instrumental noise) were mimicked by inserting the model images into actual SDSS observational fields with no apparent objects. 

After these steps we possessed ``fake'' SDSS images of the 1000 test-galaxies with known intrinsic parameters. Some typical representatives of these simulated images are shown in the lower row of Fig.~\ref{fig:mdl_sample}. For a visual comparison, the images of the corresponding real SDSS galaxies are given in the upper row of the same figure. While the actual intrinsic parameters of the upper row objects are unknown, the structure of the lower row objects is known precisely. 

For the subsequent analysis, we have used only the simulated test sample of galaxies with known parameters. Besides, we have also conducted some illustrative tests on the real SDSS sample as described in Sect.~\ref{sect:anal_obssample}.

%%======================================================================
\section{Description of the three-dimensional photometric model}
\label{sect:galmodel}

%---------------------------------------------------------------------------
\subsection{General description of the photometric model}
\label{sect:model_simple}

We use a sufficiently flexible model for describing the spatial distribution of the luminosity of a galaxy. The model galaxy is given as a superposition of its individual stellar components. Each component is approximated by an ellipsoid of rotational symmetry with a constant axial ratio $q$; its spatial density distribution follows Einasto's law
\begin{equation} l (a)=l (0)\exp \left[ -\left( {a \over ka_0}\right)^{1/N}
 \right], 
 \label{eq:einasto}
\end{equation}
where $l (0)=hL/(4\pi q a_0^3)$ is the central luminosity density and $L$ is the luminosity of the component; $a= \sqrt{r^2+z^2/q^2}$, where $r$ and $z$ are cylindrical coordinates. We make use of the harmonic mean radius $a_0$ as a good characteriser of the real extent of the component. The coefficients $h$ and $k$ are normalising parameters, dependent on the structure parameter $N$ \citep[see appendix~B of][]{Tamm:12}. The luminosity density distribution (Eq.~\ref{eq:einasto}) proposed by \citet{Einasto:65} is similar to the S\'{e}rsic law \citep{Sersic:68} for surface densities with a certain relation between corresponding structure parameters \citep{Tamm:06}.

The density distributions of all visible components are projected along the line of sight and their sum yields the surface brightness distribution of the model
\begin{equation} L(A)= 2\sum_{j}\frac{q_j}{Q_j} \int\limits_A^\infty
 \frac{l_j(a)a\,\mathrm{d}a}{(a^2-A^2)^{1/2}} , \label{eq:los_lum}
\end{equation}
where $A$ is the major semi-axis of the equidensity ellipse of the projected light distribution and $Q_j$ are their apparent axial ratios $Q^2=\cos^2i+q^2\sin^2i$. The inclination angle between the plane of the galaxy and the plane of the sky is denoted by $i$. The summation index $j$ designates each visible component.

Equation~(\ref{eq:los_lum}) gives the surface brightness of the galaxy for a given line of sight. In principle, this simple model can be further specified, by adding ring-like structures \citep[see, e.g.][]{Einasto:80} and taking account the interstellar dust inside the galaxy \citep[see, e.g.][]{Tempel:10,Tempel:11}. These details are ignored in the present analysis, since the SDSS dataset alone is not detailed enough for the inclusion of such components with sufficient confidence.

%---------------------------------------------------------------------------
\subsection{Spheroid\,+\,disc model for SDSS galaxies}
\label{sect:model_sdss}
 
Our sample of SDSS galaxies includes both early and late type galaxies. We apply simple spheroidal bulge\,+\,disc models to all galaxies independently of their morphological type. Galaxies are fitted with two Einasto's profiles (Eq.~\ref{eq:einasto}); the structural parameters $a_0$, $q$, and $N$ for the bulge and the disc are independent of each other as well as the component luminosities in $ugriz$ filters. We assume all the visible components of the galaxy to be coplanar. Although in principle, galactic components need not to be aligned, a study of a sample of 2MASS edge-on galaxies \citep{Mosenkov:10} indicates that in most cases, bulges and discs indeed are coplanar.\footnote{\citet{Mosenkov:10} determined only the position angles of the disc; the same value was ascribed to the bulges. Nevertheless, the residual images of most of the objects were found symmetrical in the bulge region in respect to the equatorial plane. This means that bulges and discs are coplanar (private communication from the authors).}

The initial guess for the centre and position angle parameters is taken from the SDSS CAS, where the surface brightness distribution of each galaxy has been approximated separately with a pure de~Vaucouleurs profile and a pure exponential disc model. During our modelling procedure, these initial parameters are adjusted separately for each filter. The same centre coordinates, position angle and inclination is assumed for the bulge and the disc of each galaxy. 

In order to keep the components realistic, we have fixed some parameter limits during modelling. For the bulges, we have set a usual lower limit of $N=2.0$ to exclude double disc-like profiles and an upper limit at $N=6.0$ to avoid unrealistically compact cores. For the discs, we have set $N=2.5$ as the upper limit. This step is similar to a multitude of 2D studies of galaxies, in which the S\'ersic index $n = 2-2.5$ is used as a watershed between galaxy discs and bulges. For the intrinsic axial ratio $q$, we have set a lower limit at 0.4 for the bulges and an upper limit at 0.3 for the discs. The latter restriction is high enough to include a vast majority of realistic discs \citep{Padilla:08, Rodriguez:13}. Note that in the present paper these limits affect only the ranges of parameters for the generated test galaxies. The used limits does not affect the modelling of the test galaxies.

When modelling real observed galaxies, the limits for model parameters should be less conservative. For example, several studies \citep{Graham:08, Gadotti:09, Laurikainen:10} have shown that the structural parameter $N$ for elliptical galaxies can be lower than 2. However, this is so only for a small fraction of galaxies. Additionally, it is known that bars alter the result of galaxy modelling if bars are not treated correctly \citep{Laurikainen:05, Gadotti:08}. Since the aim of the present paper is to model idealised test galaxies, we ignore this caveat in the following analysis.

%---------------------------------------------------------------------------
\subsection{Modelling procedure}
\label{sect:model_procedure}

We have used all the five SDSS filters ($ugriz$) in our modelling procedure. However, only the $gri$ filters were used for estimating the structural parameters since the uncertainties for $u$ and $z$ imaging are the largest. The $u$ and $z$ observations were then used for measuring the bulge and disc luminosities in these filters according to the structural model.

Correctness of the model fit was estimated by using the $\chi^2$ value, defined by
\begin{equation}
  \!\!\!\!\chi^2 \!= \frac{1}{N_\mathrm{dof}}\!\sum\limits_{\nu}\!\!\sum\limits_{x,y\in \mathrm{mask}}\!\!\!
  \frac{[f_\mathrm{obs}^\nu(x,y)-f_\mathrm{model}^\nu(x,y)]^2}{\sigma^\nu(x,y)^2},
  \label{eq:chi}
\end{equation}
where $N_\mathrm{dof}$ is the number of degrees of freedom in the fit; $f_\mathrm{obs}^\nu(x,y)$ and $f_\mathrm{model}^\nu(x,y)$ are the observed and modelled fluxes at the given pixel $(x,y)$ and index $\nu$ indicates the filter ($gri$); $\sigma(x,y)$ is the Poisson error at each pixel \citep{Howell:06}. The summation is taken over all filters $(\nu)$ and all pixels of each galaxy as defined by the corresponding mask. In the present study the number of free parameters is 16: the structural parameters $a_0$, $q$, $N$ and the $gri$ luminosities for each component and the central coordinates, inclination angle and the position angle of the galaxy. Note that Eq.~(\ref{eq:chi}) gives an equal weight to each collected photon regardless of its place of birth in the galaxy.

To minimise the $\chi^2$ we have used the downhill simplex method of Nelder and Mead from the Numerical Recipe library \citep{Press:92}. This method is quite simple and efficient in searching large parameter spaces. We have also tried other non-linear least-square fitting algorithms (e.g. Levenberg-Marquardt algorithm), but the downhill simplex method combined with simulated annealing was the most reliable one. The downhill simplex method does not require second derivatives that makes it more robust and in combination with simulated annealing it escapes the local minimums in posterior landscape.

One shortcoming of the downhill simplex method is its dependency on the initial value/guess for model parameters. In the current analysis, the initial values of the principal parameters are set according to the 2D photometry provided by the SDSS. However, when good guess values for model parameters are not available, a more time-consuming fitting methods should be used. In the most critical case of the inclination angle, we have used several different initial values across the entire realistic range. For larger sets of parameters, increasingly popular methods for that are Markov chain Monte Carlo (MCMC) ensemble sampler \citep[e.g. emcee:][]{Foreman:13} or a nested sampling tool MultiNest \citep{Feroz:13}. Alternative way is to use fast 2D fitting algorithm to estimate initial model parameters and use them for 3D fitting algorithm.

%%======================================================================
\section{Results and discussion}
\label{sect:Results}

\begin{figure}
  \centering
  \resizebox{0.5\hsize}{!}{\includegraphics{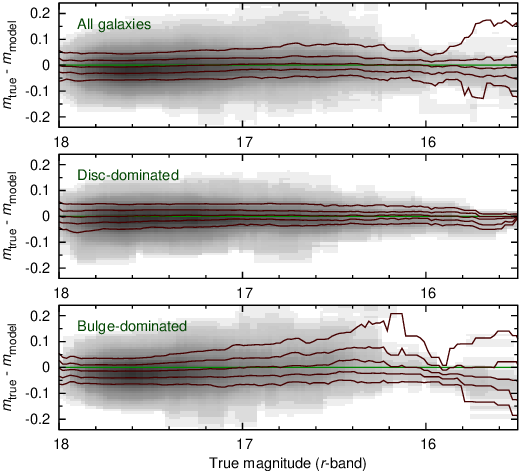}}
  \caption{Distribution of differences between the true and restored integral luminosities of all (\emph{upper panel}), disc-dominated (\emph{middle panel}), and bulge-dominated (\emph{lower panel}) galaxies, as a function of galaxy luminosity. Intensity of the \emph{shaded regions} expresses the number of objects corresponding to each distribution bin. \emph{Red solid lines} show the 0.1, 0.25, 0.5, 0.75, and 0.9 quantiles of the distribution.}
  \label{fig:magdif_model}
\end{figure}

To test the restoration accuracy of the intrinsic properties of galaxies, we have calculated differences between the properties of the original test sample galaxies and the properties estimated from the corresponding simulated images. The restoration accuracy is analysed for the integral parameters of the galaxies as well as for the bulge and disc components separately. Distinction is made also between disc-dominated and bulge-dominated galaxies on the basis of the bulge-to-disc luminosity ratio of the original galaxies in the $r$ filter: galaxies with the ratio less than 0.7 are taken to be disc-dominated and galaxies with the ratio larger than 1.3 are taken to be bulge-dominated. 

%---------------------------------------------------------------------------
\subsection{Uncertainties of the intrinsic luminosities and colours}
\label{sect:res_luminosity}

%\subsubsection{Total luminosities of galaxies}

%\subsubsection{Luminosities of bulges and discs}
\begin{figure}
  \centering
  \resizebox{0.5\hsize}{!}{\includegraphics{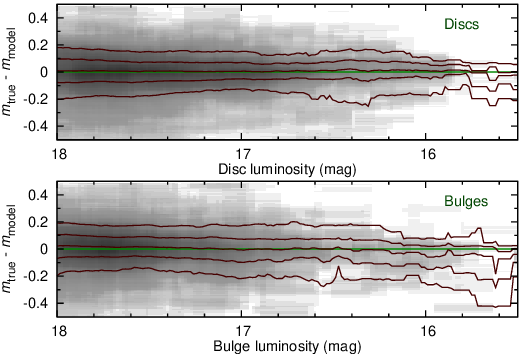}}
  \caption{Distribution of differences between the true and restored integral luminosities of the disc (\emph{upper panel}) and bulge (\emph{lower panel}) components of the galaxies as a function of the component luminosity. \emph{Red solid lines} show the 0.1, 0.25, 0.5, 0.75, and 0.9 quantiles of the distribution.}
  \label{fig:magdif_discbulge}
\end{figure}

\begin{figure}
  \centering
  \resizebox{0.5\hsize}{!}{\includegraphics{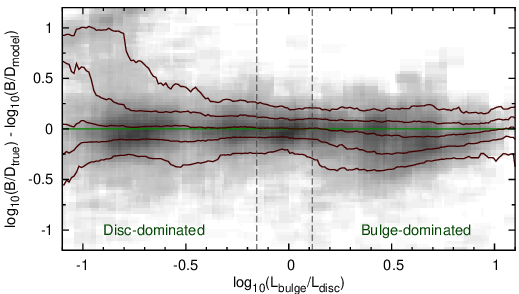}}
  \caption{Distribution of differences between the true and restored bulge-to-disc luminosity ratios (B/D) as a function of B/D. \emph{Red solid lines} show the 0.1, 0.25, 0.5, 0.75, and 0.9 quantiles of the distribution. Vertical \emph{dashed lines} separate the regions of disc-dominated and bulge-dominated galaxies as defined in this study. For galaxies with B/D less than 0.2, bulge component is poorly determined, yielding large uncertainties. On an average, no general systematic shift of the components fraction is introduced.}
  \label{fig:dif_bdratio}
\end{figure}

Differences between the original and restored integral $r$-band luminosities of the galaxies are shown in Fig.~\ref{fig:magdif_model}; the qualitative results for the other SDSS filters are similar. As indicated by the quantile lines in the figure, differences are below 0.05~mag for most of the galaxies, almost independently of galaxy luminosity. The errors are similar to the original SDSS galaxy catalogue errors of the total luminosities of galaxies, thus indicating the dominance of the photometric uncertainties of the images over the accuracy of the modelling or measuring technique. For brighter bulge-dominated galaxies, the errors become slightly larger. Luminosities are systematically slightly underestimated, probably related to the uncertainties of other parameters of these galaxies. However, the number of bright bulge-dominated galaxies is relatively small, hence, this behaviour might be just a statistical fluctuation.

Figure~\ref{fig:magdif_discbulge} shows differences between the true and modelled integral luminosities for disc and bulge components of galaxies. In general, the differences are somewhat larger (within 0.2~mag) than for the whole galaxies, which is a natural result of the degeneracy between bulge and disc components. In most cases, the bulge and disc luminosity determination errors compensate each other and the total luminosity is conserved. For bright bulges, an increasing luminosity underestimation can be noted, probably for similar reasons as in the case of bright bulge-dominated galaxies.

Figure~\ref{fig:dif_bdratio} shows differences between the modelled and true B/D as a function of true B/D. It is seen that for $\mathrm{B/D} > 0.3$, the errors are independent of the initial B/D and typically stay within $\pm$40\%. For galaxies with very small B/D (close to 0.2), the bulge component is rather inaccurately recovered, yielding large uncertainties in B/D determination. However, the total luminosity of the galaxy remains almost unaffected. On the other hand, the disc luminosity is estimated accurately also in the case of high B/D. Despite the generally low accuracy of the restored B/D, it is important to note that on an average, no systematic shift of the balance between bulges and discs is artificially introduced.

%\subsubsection{Luminosities in inner and outer parts of a galaxy}
\begin{figure}
  \centering
  \resizebox{0.5\hsize}{!}{\includegraphics{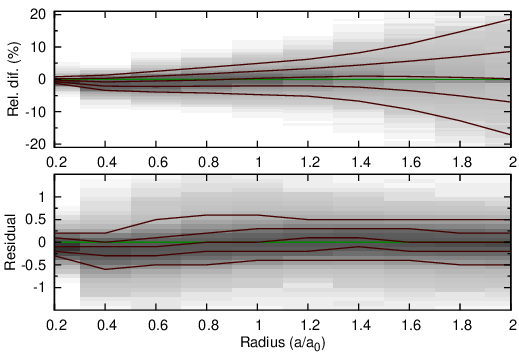}}
  \caption{Luminosity difference distribution of true and modelled galaxies, as measured within thin concentric rings as a function of ring radius. \emph{Upper panel:} luminosity differences divided by the luminosity (per cent). \emph{Lower panel:} residual values in the SDSS standard units, nanomaggies. Ring radii are presented as a fraction of the harmonic mean radius of the given galaxy. \emph{Red solid lines} show the 0.1, 0.25, 0.5, 0.75, and 0.9 quantiles of the distribution.}
  \label{fig:rdif_model}
\end{figure}

The relative and absolute differences between the true and modelled luminosities of galaxies as a function of distance from galaxy centre are presented in Fig.~\ref{fig:rdif_model}. The luminosities are measured within thin concentric rings. The ring radii are presented as fractions of the harmonic mean radii $a_0$ of the galaxies (more precisely, as fractions of the radii of the larger component of the given galaxy). For inner regions, the modelling accuracy stays well within the 5\% limits. As the surface brightness drops and noise starts to dominate in the outer regions, the models become less accurate as expected, but no systematic trend is introduced. The lower panel in Fig.~\ref{fig:rdif_model} shows the residuals measured from the simulated SDSS images inside concentric circles after the subtraction of the model. The residuals are radius-independent, because during our model fitting procedure, the inner and outer luminosities are considered with equal weight (see Sect.~\ref{sect:model_procedure}).

%\subsubsection{Colour indexes}

\begin{figure}
  \centering
  \resizebox{0.5\hsize}{!}{\includegraphics{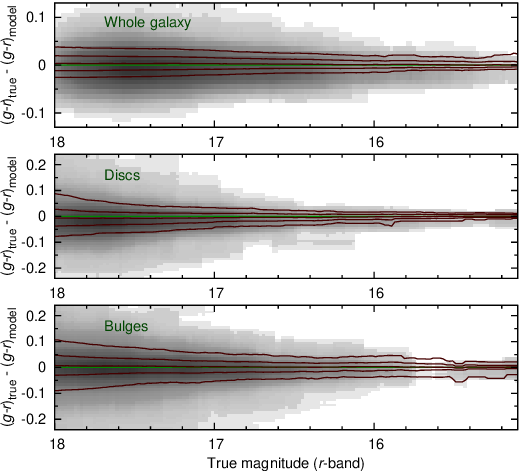}}
  \caption{Distribution of differences between true and modelled integral $g-r$ colours for whole galaxies (\emph{upper panel}), for disc components (\emph{middle panel}), and for bulge components (\emph{lower panel}) as a function of luminosity. \emph{Red solid lines} show the 0.1, 0.25, 0.5, 0.75, and 0.9 quantiles of the distribution. In general, differences are smaller than for integral luminosities.}
  \label{fig:colind}
\end{figure}

Differences between the original and modelled $g-r$ colour indices are presented in Fig.~\ref{fig:colind}. For all the other colour indices, the results are similar. The errors remain within 0.03~mag for most of the galaxies, being smaller than the errors of integral luminosities. This is mainly caused by the fact that the fitted model galaxy has the same structural parameters for each filter, thus the systematic errors in the structure recovery are also similar for each filter, leading to quantitatively similar misestimation of the luminosity in each filter and a correspondingly smaller spread in the colour indices \citep[the same conclusion is reached in][]{DeLooze:14}. As expected, colours are estimated more accurately for brighter galaxies and galactic components. At the faint end of the sample, the errors increase up to 0.1~mag. Similarly to integral luminosities, colours of disc components are recovered with higher precision than colours of bulge components.

%---------------------------------------------------------------------------
\subsection{Uncertainties of the inclination angles of galaxies}
\label{sect:res_inclination}

\begin{figure}
  \centering
  \resizebox{0.5\hsize}{!}{\includegraphics{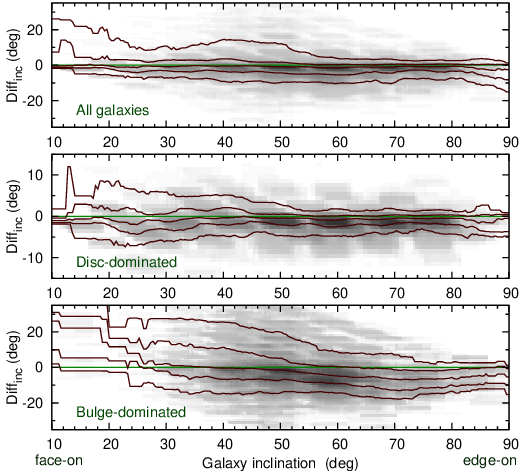}}
  \caption{Distribution of differences between the true and restored inclination angles of galaxies for all (\emph{upper panel}), disc-dominated (\emph{middle panel}), and bulge-dominated (\emph{lower panel}) galaxies, presented in degrees. The slight decrease of differences for high inclinations is due the fact that inclination angle cannot be larger than 90$^\circ$. \emph{Red solid lines} show the 0.1, 0.25, 0.5, 0.75, and 0.9 quantiles of the distribution.}
  \label{fig:incdif}
\end{figure}

The main advantage of 3D galactic models lies in their direct yield of the spatial density distribution, including the inclination angle, thus making them useful for dynamical analyses and studies of the large-scale alignment of galactic structures. However, this advantage comes with a considerable sacrifice of computational time, thus the accuracy of inclination angle estimation is crucial in deciding whether or not 3D modelling is worth attempting.

To measure the inclination angle of a galaxy, we run the modelling code with ten different initial guesses for the inclination angle between $0\degr$ (face-on galaxies) and $90\degr$ (edge-on galaxies). The modelling was done separately for the $gri$ images, providing three independent values for the inclination angle. For a majority of disc-dominated galaxies, these three values were very close. In the final modelling step, the $gri$ filter data were used simultaneously and the best estimate for the inclination angle was determined by minimising the $\chi^2$.

\begin{figure}
  \centering
  \resizebox{0.5\hsize}{!}{\includegraphics{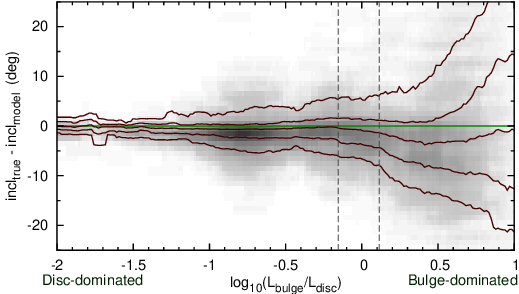}}
  \caption{Distribution of differences between the true and restored inclination angles of galaxies as a function of B/D. \emph{Vertical dashed lines} separate the regions of disc-dominated and bulge-dominated galaxies as defined in this study. The errors are mostly below 5$^\circ$ for disc-dominated galaxies but worsens while moving toward bulge-dominated (i.e. elliptical) galaxies. At $\mathrm{B/D} > 6$ the inclination angle cannot be determined. \emph{Red solid lines} show the 0.1, 0.25, 0.5, 0.75, and 0.9 quantiles of the distribution.}
  \label{fig:incdif_bdratio}
\end{figure}

Differences between the true and modelled inclination angles are shown in Fig.~\ref{fig:incdif}. For disc-dominated galaxies, the errors are below 5 degrees in most cases, being slightly lower for more edge-on galaxies. The slight decrease of differences for large inclinations is due the fact that inclination angle cannot be larger than 90$^\circ$. 
Since the inclination angle of a galaxy is related to its disc rather than bulge, the inclination angle restoration accuracy correlates with the B/D of the galaxy, as seen in Fig.~\ref{fig:incdif_bdratio}. Differences between the true and restored inclinations start to increase rapidly, when the B/D becomes larger than 2--3. At $\mathrm{B/D} > 6$, we can consider the galaxies as pure ellipticals and the inclination angle becomes indeterminable. According to our preliminary results for nearly 500000 SDSS galaxies, the criterion $\mathrm{B/D} < 6$ includes roughly 75\% of galaxies in the nearby Universe; in the more distant Universe ($z > 0.5$), irregular and peculiar morphologies start to dominate, but the regular spiral galaxies still form a substantial fraction \citep{DelgadoSerrano2012}.

Note that in principle, the inclination angles of real elliptical galaxies cannot be restored (and are difficult to define!) from images since these objects are triaxial ellipsoids, but if a galaxy contains a significant disc component, the inclination angle can be determined from it. Nevertheless, the influence of the bulge component to the inclination angle determination is still present via the uncertainty of the bulge-disc decomposition.

Based on this test, we can conclude that inclination angles can be restored sufficiently accurately for spiral and S0 galaxies and the additionally required CPU time for 3D modelling is justified, when the inclination angle determination is important. In \citet{Tempel:12a} it is shown that 3D modelling can give statistically slightly better inclination angles than a simple 2D modelling.

Since our method to model galaxies depends slightly on the initial parameters, due to the degeneracy between various parameters, we cannot find all the parameters directly. The degeneracy is strongest for the inclination angles of galaxies, which are degenerate with the thickness/ellipticity of the galaxy. However, it is major problem only for bulge-dominated galaxies.

The reader may notice the lack of face-on bulge-dominated galaxies in our sample. Our sample of idealised model galaxies is constructed on the basis on actual objects and the inclination angle is derived assuming axially symmetric galaxies. In reality, bulges and elliptical galaxies are triaxial ellipsoids, yielding no circularly symmetric projections which could be interpreted as a face-on configuration.

%--------------------------------------------------------------------------
\subsection{Uncertainties of the structural parameters}
\label{sect:res_struct_pars}

\begin{figure}
  \centering
  \resizebox{0.5\hsize}{!}{\includegraphics{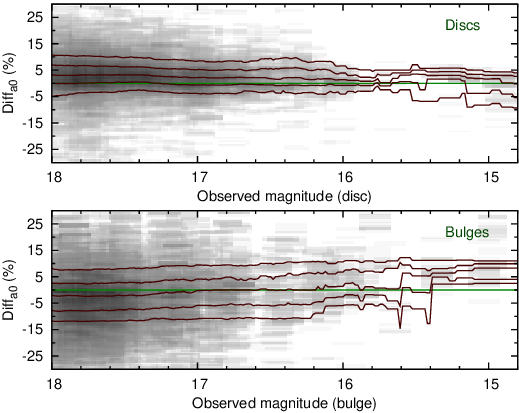}}
  \caption{Distribution of differences (per cent) between the original and restored harmonic mean radius $a_0$ for disc (\emph{upper panel}) and for bulge (\emph{lower panel}) component as a function of the component luminosity. \emph{Red solid lines} show the 0.1, 0.25, 0.5, 0.75, and 0.9 quantiles of the distribution.}
  \label{fig:dif_a0}
\end{figure}

Figure~\ref{fig:dif_a0} shows relative differences between the true and modelled harmonic mean radius $a_0$ as a function of component luminosity. The figure shows that for discs, the difference is almost insensitive to the luminosity. Since luminosity correlates with radius, the difference is also insensitive to the disc radius. It is noticeable, that the modelled discs are systematically smaller by about a few per cent. In general, for most of the galaxies, the disc radius is restored with accuracy higher than 10 per cent.

For bulges (lower panel in Fig.~\ref{fig:dif_a0}), the accuracy of modelled radius $a_0$ is worser than for discs. For brighter bulges, the modelled radius is five to ten per cent below the true value. This effect is probably responsible for the underestimation of the luminosities of bright bulges.

\begin{figure}
  \centering
  \resizebox{0.5\hsize}{!}{\includegraphics{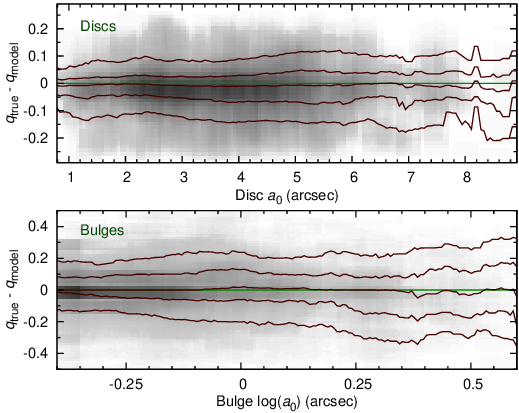}}
  \caption{Difference between true and modelled axial ratio $q$ for disc (\emph{upper panel}) and bulge (\emph{lower panel}) components as a function of the component radius $a_0$. \emph{Red solid lines} show the 0.1, 0.25, 0.5, 0.75, and 0.9 quantiles of the distribution.}
  \label{fig:dif_q}
\end{figure}

Figure~\ref{fig:dif_q} shows differences between the true and modelled axial ratios $q$ of the galactic components as a function of the radius of the component. Considering the upper limit for the disc axial ratio $q=0.3$ in the model, the axial ratio accuracy $\pm 0.05$ for about half of the test galaxies is satisfactory. For about 20\% of the discs (outside the 0.1 and 0.9 quantile lines), the model estimate totally misses the true value, being off by more than 0.1 (i.e., one third of the allowed range). For the rest, the model accuracy is low, but may provide some statistically reliable results in large surveys. For the bulges, differences seem even larger, but the allowed range for $q$ is also considerably larger (0.4--1.0). Figure~\ref{fig:dif_q} also shows that the accuracy of $q$ determination does not depend on the size of the component.

\begin{figure}
  \centering
  \resizebox{0.5\hsize}{!}{\includegraphics{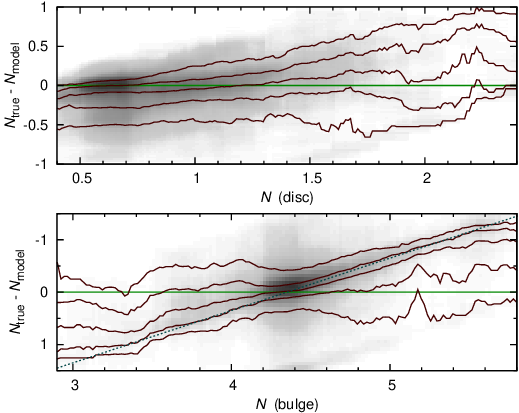}}
  \caption{Differences between the true and modelled structural parameter $N$ for disc (\emph{upper panel}) and bulge (\emph{lower panel}) components as a function of $N$. \emph{Red solid lines} show the 0.1, 0.25, 0.5, 0.75, and 0.9 quantiles of the distribution. The tilt of the distributions is caused by the lower and upper limits for $N$. The \emph{dotted} straight line in the \emph{lower} panel corresponds to the restored parameter $N=4.35$.}
  \label{fig:dif_n}
\end{figure}

Figure~\ref{fig:dif_n} shows differences between the true and modelled values for the structural parameter $N$. The decreasing trend of both distributions is caused by the lower and upper limits set for $N$ during modelling. Close to these limits, the difference can only be positive or negative, respectively. As expected, the accuracy of $N$ is rather low, staying within about $\pm 0.5$ for most of the cases.

\begin{figure*}
  \centering
  \resizebox{\hsize}{!}{\includegraphics{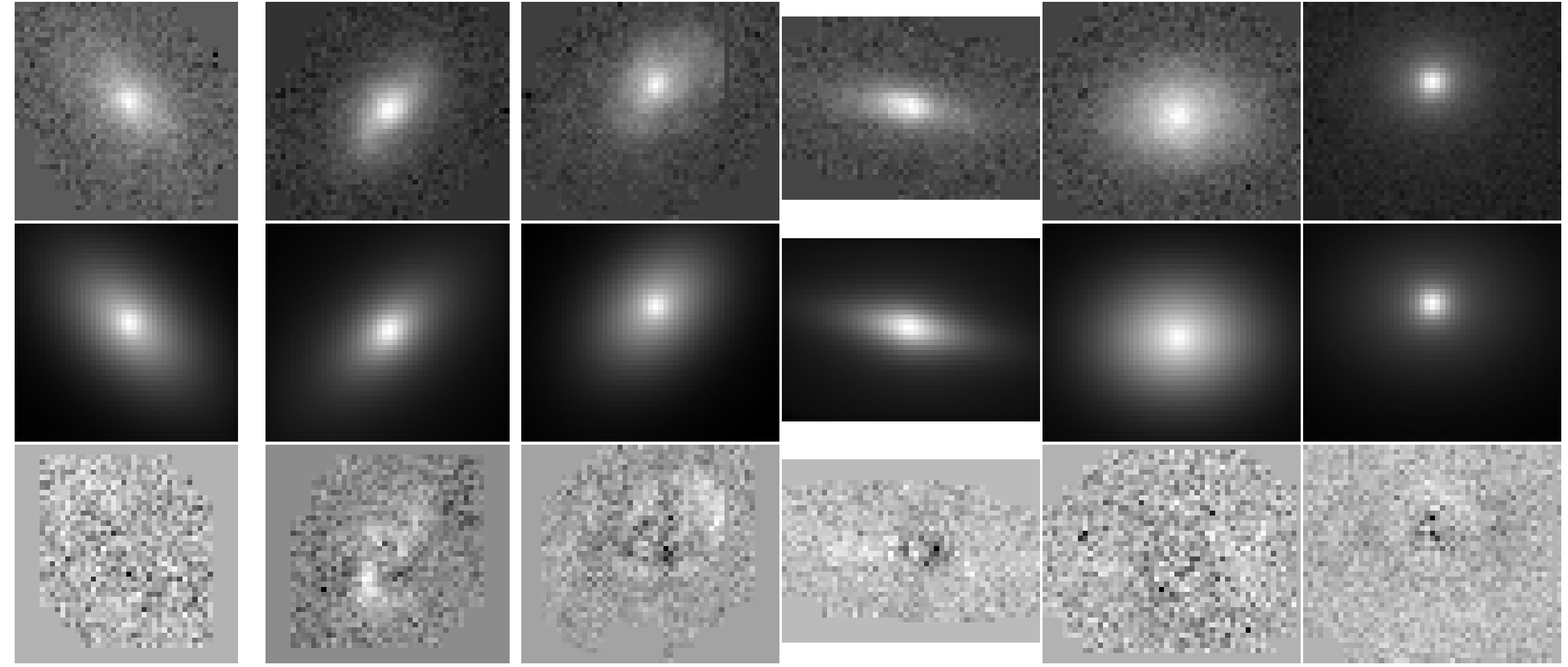}}
  \caption{Examples of modelling real SDSS galaxies. \emph{Upper row} shows the original observations, \emph{middle} row shows the PSF-convolved model galaxies, and \emph{lower row} shows the residual images.}
  \label{fig:mdl_model_sky}
\end{figure*}

For the bulges (lower panel in Fig.~\ref{fig:dif_n}), we can note clustering around the value $N\approx 4.3$, which corresponds to $N\approx 4$ for the S\'{e}rsic 2D distribution \citep{Dhar:10}. On one hand, this is known to be the most typical value for bulges. On the other hand, we have used it as the initial guess value for the bulge components. The dotted line plotted for the restored value $N=4.35$ indicates that the model has usually found a solution close to the initial value. This suggests that in most cases, it is reasonable to fix the bulge $N$ at 4.3 during modelling and let the parameter free only in the case of the appearance of significant residuals, as has been recently done with 2D models \citep{Simard:11,Lackner:12,Meert:15}.

%%======================================================================
\section{Modelling the real SDSS images}
\label{sect:anal_obssample}

\begin{figure}
  \centering
  \resizebox{0.5\hsize}{!}{\includegraphics{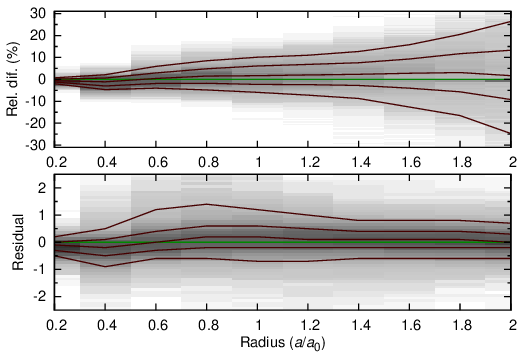}}
  \caption{Differences between the observed and modelled luminosities of galaxies, measured inside concentric rings as a function of ring radius. \emph{Upper panel:} luminosity differences divided by the luminosity (per cent). \emph{Lower panel:} residual values in the SDSS standard units, nanomaggies. The ring radii are presented as a fraction of the harmonic mean radius of the given galaxy. \emph{Red solid lines} show the 0.1, 0.25, 0.5, 0.75, and 0.9 quantiles of the distribution.}
  \label{fig:rdif_modelobs}
\end{figure}

Above, the reliability of modelling idealistic test galaxies was presented. The actual galaxies are usually far from such simplified objects. Instead, they contain spiral structure, rings, asymmetries, dust lanes, varying inclination and position angles, etc. So let us take a look at the reproduction of some of the actual galaxies from the original SDSS images during the creation of the test galaxy sample, which was described in Sect.~\ref{sect:sdss_data}.

Figure~\ref{fig:mdl_model_sky} shows some examples of the original SDSS images, the model galaxies and the residual images. As expected, the spiral structure is still present on the residual images of disc galaxies as well as some asymmetries, which is expected since the model galaxies are axisymmetric. However, no luminosity gradients can be detected, thus the axisymmetric distributions have been recovered correctly.

The upper panel in Fig.~\ref{fig:rdif_modelobs} shows the distribution of relative differences between the true and modelled luminosities within thin concentric rings as a function of ring radius. For this figure, galaxy sizes are normalised according to their harmonic mean radii $a_0$. Comparing to the models of test galaxies (Fig.~\ref{fig:rdif_model}) the differences are now slightly larger due to the substructure in actual galaxies.

Lower panel in Fig.~\ref{fig:rdif_model} shows the distribution of luminosity differences in SDSS standard units (nano\-maggies), as measured from the residual images with\-in concentric rings. The differences remain within 1 nanomaggies and do not depend significantly on the distance from the centre of the galaxy.

\section{Conclusions}

In this paper we have analysed the reliability of 3D structural models. For the analysis, artificial SDSS images of idealistic model galaxies were created, extending down to the $r$-band limiting magnitude 17.77, and the accuracy of the recovered parameters was estimated. Within the given luminosity limits, the integral luminosities and colours were recovered with high precision ($\pm$0.05~mag). The luminosities of individual components were somewhat less accurate ($\pm$0.2~mag). The resultant B/D luminosity ratio is within $\pm$40\% of the initial value, but on an average, no systematic shift is introduced, thus the method can be used for measuring the statistical fractions of bulges (spheroids) and discs in large samples.

Given that many early-type galaxies contain detectable disc structures, galaxy inclination angles can be estimated for a broad range of morphologies. As expected, the inclination angle estimates are better for disc-do\-mi\-nated galaxies, with the errors remaining mostly below 5$^\circ$ for galaxies with $\mathrm{B/D} < 2$ and below $15^\circ$ for galaxies with $\mathrm{B/D} < 6$; the latter criterion involves the majority of actual galaxies (roughly 75\%) in the nearby Universe. Errors of the recovered sizes of the galactic components are less than 10\% in most cases. Axial ratios and the parameter $N$ of Einasto's distribution (similar to the S\'{e}rsic index) are relatively inaccurate, but can provide statistical estimates for studies of large galactic samples. In general, models of disc components are more accurate than models of spheroidal components, which is a natural effect of the spatial geometry.

We can conclude that especially for statistical studies, 3D modelling is worth the extra computation time needed, allowing us to estimate parameters (e.g. inclination angles) which are not directly accessed with 2D methods. For example, if applied on large samples of galaxies, such a modelling can reveal correlations between inclination angles and the large-scale structure of the Universe \citep{Tempel:12a}. Note that such an analysis can be done also by 2D modelling, but as discussed in \citet{Zhang:13}, a careful real space alignment analysis is necessary to detect the weak alignment signal.

The structural parameters, if determined from images only, are not always reliable for any given object: mostly because of the degeneracies between the galaxy structural parameters. However, unlike the case of a 2D approximation of the surface density, the derived parameters are essentially inclination-independent. Moreover, if data about the intrinsic kinematics of a galaxy are available, the 3D modelling will become a powerful tool for measuring galactic structures and masses \citep{Tamm:12}.

We have used idealised test galaxies in the presented study. It is natural that the properties of real galaxies are more difficult to determine because of non-symmetries and additional components. An analysis of 3D modelling accuracy of real well-studied nearby galaxies is the topic of a forthcoming paper.

\normalem
\begin{acknowledgements}
We thank the referee for constructive remarks and E.~Saar and J.~Pelt for helpful discussion and suggestions. This work was supported by the Estonian Science Foundation projects IUT26-2, IUT40-2. We acknowledge the support by the Centre of Excellence of Dark Matter in (Astro)particle Physics and Cosmology (TK120). Some of the preliminary data analysis have been done with an interactive graphical tool TOPCAT (http://www.starlink.ac.uk/topcat/). Time consuming computations were carried out at the High Performance Computing Centre, University of Tartu. We are pleased to thank the SDSS-III Team for the publicly available data releases. Funding for SDSS-III has been provided by the Alfred P. Sloan Foundation, the Participating Institutions, the National Science Foundation, and the U.S. Department of Energy Office of Science. The SDSS-III web site is http://www.sdss3.org/.  

\end{acknowledgements}

%\bibliographystyle{raa}
%\bibliography{galbib.bib}

\begin{thebibliography}{45}
\providecommand{\natexlab}[1]{#1}
\providecommand{\selectlanguage}[1]{\relax}

\bibitem[{{Aihara} et~al.(2011){Aihara}, {Allende Prieto}, {An}
  et~al.}]{Aihara:11}
{Aihara}, H., {Allende Prieto}, C., {An}, D., et~al. 2011, ApJS, 193, 29

\bibitem[{{Arnold} et~al.(2014){Arnold}, {Romanowsky}, {Brodie}
  et~al.}]{Arnold:14}
{Arnold}, J.~A., {Romanowsky}, A.~J., {Brodie}, J.~P., et~al. 2014, \apj, 791,
  80

\bibitem[{{Bak} \& {Statler}(2000)}]{BakStat00}
{Bak}, J., \& {Statler}, T.~S. 2000, \aj, 120, 110

\bibitem[{{Bertola} et~al.(1991){Bertola}, {Bettoni}, {Danziger}
  et~al.}]{Bertola91}
{Bertola}, F., {Bettoni}, D., {Danziger}, J., et~al. 1991, \apj, 373, 369

\bibitem[{{De Looze} et~al.(2014){De Looze}, {Fritz}, {Baes}
  et~al.}]{DeLooze:14}
{De Looze}, I., {Fritz}, J., {Baes}, M., et~al. 2014, \aap, 571, A69

\bibitem[{{de Souza} et~al.(2004){de Souza}, {Gadotti}, \& {dos
  Anjos}}]{deSouza:04}
{de Souza}, R.~E., {Gadotti}, D.~A., \& {dos Anjos}, S. 2004, ApJS, 153, 411

\bibitem[{{Delgado-Serrano} et~al.(2010){Delgado-Serrano}, {Hammer}, {Yang}
  et~al.}]{DelgadoSerrano2012}
{Delgado-Serrano}, R., {Hammer}, F., {Yang}, Y.~B., et~al. 2010, \aap, 509, A78

\bibitem[{{Dhar} \& {Williams}(2010)}]{Dhar:10}
{Dhar}, B.~K., \& {Williams}, L.~L.~R. 2010, MNRAS, 405, 340

\bibitem[{{Einasto}(1965)}]{Einasto:65}
{Einasto}, J. 1965, Tartu Astr. Obs. Teated, 17

\bibitem[{{Einasto} et~al.(1980){Einasto}, {Tenjes}, {Barabanov}, \&
  {Zasov}}]{Einasto:80}
{Einasto}, J., {Tenjes}, P., {Barabanov}, A.~V., \& {Zasov}, A.~V. 1980,
  Ap\&SS, 67, 31

\bibitem[{{Feroz} et~al.(2013){Feroz}, {Hobson}, {Cameron}, \&
  {Pettitt}}]{Feroz:13}
{Feroz}, F., {Hobson}, M.~P., {Cameron}, E., \& {Pettitt}, A.~N. 2013,
  arXiv:1306.2144

\bibitem[{{Foreman-Mackey} et~al.(2013){Foreman-Mackey}, {Hogg}, {Lang}, \&
  {Goodman}}]{Foreman:13}
{Foreman-Mackey}, D., {Hogg}, D.~W., {Lang}, D., \& {Goodman}, J. 2013, \pasp,
  125, 306

\bibitem[{{Gadotti}(2008)}]{Gadotti:08}
{Gadotti}, D.~A. 2008, \mnras, 384, 420

\bibitem[{{Gadotti}(2009)}]{Gadotti:09}
{Gadotti}, D.~A. 2009, \mnras, 393, 1531

\bibitem[{{Graham} \& {Worley}(2008)}]{Graham:08}
{Graham}, A.~W., \& {Worley}, C.~C. 2008, \mnras, 388, 1708

\bibitem[{{Howell}(2006)}]{Howell:06}
{Howell}, S.~B. 2006, {Handbook of CCD astronomy} (Cambridge University Press)

\bibitem[{{Kautsch} et~al.(2006){Kautsch}, {Grebel}, {Barazza}, \&
  {Gallagher}}]{Kautsch:06}
{Kautsch}, S.~J., {Grebel}, E.~K., {Barazza}, F.~D., \& {Gallagher}, J.~S., III
  2006, \aap, 445, 765

\bibitem[{{Kipper} et~al.(2012){Kipper}, {Tempel}, \& {Tamm}}]{Kipper:12}
{Kipper}, R., {Tempel}, E., \& {Tamm}, A. 2012, Baltic Astronomy, 21, 523

\bibitem[{{Krajnovi{\'c}} et~al.(2008){Krajnovi{\'c}}, {Bacon}, {Cappellari}
  et~al.}]{Krajnovic:08}
{Krajnovi{\'c}}, D., {Bacon}, R., {Cappellari}, M., et~al. 2008, \mnras, 390,
  93

\bibitem[{{Krajnovi{\'c}} et~al.(2013){Krajnovi{\'c}}, {Alatalo}, {Blitz}
  et~al.}]{Krajnovic:13}
{Krajnovi{\'c}}, D., {Alatalo}, K., {Blitz}, L., et~al. 2013, \mnras, 432, 1768

\bibitem[{{Lackner} \& {Gunn}(2012)}]{Lackner:12}
{Lackner}, C.~N., \& {Gunn}, J.~E. 2012, \mnras, 421, 2277

\bibitem[{{Laurikainen} et~al.(2005){Laurikainen}, {Salo}, \&
  {Buta}}]{Laurikainen:05}
{Laurikainen}, E., {Salo}, H., \& {Buta}, R. 2005, \mnras, 362, 1319

\bibitem[{{Laurikainen} et~al.(2010){Laurikainen}, {Salo}, {Buta}, {Knapen}, \&
  {Comer{\'o}n}}]{Laurikainen:10}
{Laurikainen}, E., {Salo}, H., {Buta}, R., {Knapen}, J.~H., \& {Comer{\'o}n},
  S. 2010, \mnras, 405, 1089

\bibitem[{{Meert} et~al.(2015){Meert}, {Vikram}, \& {Bernardi}}]{Meert:15}
{Meert}, A., {Vikram}, V., \& {Bernardi}, M. 2015, \mnras, 446, 3943

\bibitem[{{Mendel} et~al.(2014){Mendel}, {Simard}, {Palmer}, {Ellison}, \&
  {Patton}}]{Mendel:14}
{Mendel}, J.~T., {Simard}, L., {Palmer}, M., {Ellison}, S.~L., \& {Patton},
  D.~R. 2014, \apjs, 210, 3

\bibitem[{{Mosenkov}(2014)}]{Mosenkov:14}
{Mosenkov}, A.~V. 2014, Astrophysical Bulletin, 69, 99

\bibitem[{{Mosenkov} et~al.(2010){Mosenkov}, {Sotnikova}, \&
  {Reshetnikov}}]{Mosenkov:10}
{Mosenkov}, A.~V., {Sotnikova}, N.~Y., \& {Reshetnikov}, V.~P. 2010, \mnras,
  401, 559

\bibitem[{{Padilla} \& {Strauss}(2008)}]{Padilla:08}
{Padilla}, N.~D., \& {Strauss}, M.~A. 2008, \mnras, 388, 1321

\bibitem[{{Peng} et~al.(2010){Peng}, {Ho}, {Impey}, \& {Rix}}]{Peng:10}
{Peng}, C.~Y., {Ho}, L.~C., {Impey}, C.~D., \& {Rix}, H.-W. 2010, AJ, 139, 2097

\bibitem[{{Press} et~al.(1992){Press}, {Teukolsky}, {Vetterling}, \&
  {Flannery}}]{Press:92}
{Press}, W.~H., {Teukolsky}, S.~A., {Vetterling}, W.~T., \& {Flannery}, B.~P.
  1992, {Numerical recipes in FORTRAN. The art of scientific computing}

\bibitem[{{Rodr{\'{\i}}guez} \& {Padilla}(2013)}]{Rodriguez:13}
{Rodr{\'{\i}}guez}, S., \& {Padilla}, N.~D. 2013, \mnras, 434, 2153

\bibitem[{{S\'{e}rsic}(1968)}]{Sersic:68}
{S\'{e}rsic}, J.~L. 1968 (Cordoba, Argentina: Observatorio Astronomico)

\bibitem[{{Simard} et~al.(2002){Simard}, {Willmer}, {Vogt} et~al.}]{Simard:02}
{Simard}, L., {Willmer}, C.~N.~A., {Vogt}, N.~P., et~al. 2002, ApJS, 142, 1

\bibitem[{{Simard} et~al.(2011){Simard}, {Mendel}, {Patton}, {Ellison}, \&
  {McConnachie}}]{Simard:11}
{Simard}, L., {Mendel}, J.~T., {Patton}, D.~R., {Ellison}, S.~L., \&
  {McConnachie}, A.~W. 2011, ApJS, 196, 11

\bibitem[{{Strauss} et~al.(2002){Strauss}, {Weinberg}, {Lupton}
  et~al.}]{Strauss:02}
{Strauss}, M.~A., {Weinberg}, D.~H., {Lupton}, R.~H., et~al. 2002, AJ, 124,
  1810

\bibitem[{{Tamm} \& {Tenjes}(2006)}]{Tamm:06}
{Tamm}, A., \& {Tenjes}, P. 2006, A\&A, 449, 67

\bibitem[{{Tamm} et~al.(2012){Tamm}, {Tempel}, {Tenjes}, {Tihhonova}, \&
  {Tuvikene}}]{Tamm:12}
{Tamm}, A., {Tempel}, E., {Tenjes}, P., {Tihhonova}, O., \& {Tuvikene}, T.
  2012, A\&A, 546, A4

\bibitem[{{Tempel} et~al.(2010){Tempel}, {Tamm}, \& {Tenjes}}]{Tempel:10}
{Tempel}, E., {Tamm}, A., \& {Tenjes}, P. 2010, A\&A, 509, A91

\bibitem[{{Tempel} et~al.(2011){Tempel}, {Tuvikene}, {Tamm}, \&
  {Tenjes}}]{Tempel:11}
{Tempel}, E., {Tuvikene}, T., {Tamm}, A., \& {Tenjes}, P. 2011, A\&A, 526, A155

\bibitem[{{Tempel} et~al.(2013){Tempel}, {Stoica}, \& {Saar}}]{Tempel:12a}
{Tempel}, E., {Stoica}, R.~S., \& {Saar}, E. 2013, \mnras, 428, 1827

\bibitem[{{Tenjes} et~al.(1993){Tenjes}, {Busarello}, {Longo}, \&
  {Zaggia}}]{Tenjes93}
{Tenjes}, P., {Busarello}, G., {Longo}, G., \& {Zaggia}, S. 1993, A\&A, 275, 61

\bibitem[{{van den Bosch} \& {van de Ven}(2009)}]{vandeven09}
{van den Bosch}, R.~C.~E., \& {van de Ven}, G. 2009, MNRAS, 398, 1117

\bibitem[{{Vikram} et~al.(2010){Vikram}, {Wadadekar}, {Kembhavi}, \&
  {Vijayagovindan}}]{Vikram:10}
{Vikram}, V., {Wadadekar}, Y., {Kembhavi}, A.~K., \& {Vijayagovindan}, G.~V.
  2010, \mnras, 409, 1379

\bibitem[{{York} et~al.(2000){York}, {Adelman}, {Anderson} et~al.}]{York:00}
{York}, D.~G., {Adelman}, J., {Anderson}, J.~E., Jr., et~al. 2000, \aj, 120,
  1579

\bibitem[{{Zhang} et~al.(2013){Zhang}, {Yang}, {Wang} et~al.}]{Zhang:13}
{Zhang}, Y., {Yang}, X., {Wang}, H., et~al. 2013, \apj, 779, 160

\end{thebibliography}

\end{document}